\documentstyle[psfig]{l-aa}
\begin{document}
   \thesaurus{12         
              (11.03.4 Abell 370;  
               11.05.2;  
               11.19.3;  
               12.03.3;  
               12.07.1)} 

\title{Lensed galaxies in Abell 370}
\subtitle{II. The ultraviolet properties of arclets and the star
formation rate at high redshift
\thanks{Based on observations with the NASA/ESA {\it Hubble
Space Telescope\/} obtained from the data archive at the Space Telescope
Science Institute, USA}
}

   \author{
J. B\'ezecourt\inst{1} \and G. Soucail\inst{2} \and R.S.
Ellis\inst{3} \and J-P. Kneib\inst{2}
   }

   \offprints{J. B\'ezecourt, bezecour@astro.rug.nl}

   \institute{Kapteyn Institute, Postbus 800, 9700 AV Groningen, The Netherlands
   \and Observatoire Midi-Pyr\'en\'ees, Laboratoire d'Astrophysique,
    UMR 5572, 14 Avenue E. Belin, F-31400 Toulouse, France
   \and Institute of Astronomy, Madingley Road, Cambridge CB3 0HA, UK}

\date{Received , Accepted }

\maketitle


\begin{abstract}

We discuss the statistical properties of faint background galaxies
detected in a deep Hubble Space Telescope ultraviolet (F336W) Wide
Field Planetary Camera 2 image of the lensing cluster Abell 370
($z$=0.37). By combining modelled field galaxy counts at this
wavelength with a detailed mass model for the cluster, we develop
techniques for predicting the expected redshift distribution of
lensed sources and find the bulk of the sources should be
moderately magnified sources whose redshifts lie between 0.5 and
2. We compare these predictions with redshift estimates derived
from the lens inversion method which utilises the shapes of
individual arclets in the context of a known field ellipticity
distribution. This allows us to infer the comoving star formation
rate for the range $0.5<z<2$ where there is currently little
spectroscopic data. We discuss our results in the context of
contemporary pictures for the history of star formation.

\end{abstract}

\section{Introduction}

A detailed understanding of galaxy formation remains a major
question of modern cosmology. An important observational indicator
is the history and physics of star formation resulting from the
induced perturbations of gas clouds in the rapidly evolving
gravitational potential. The concept of a single epoch for galaxy
formation associated with "monolithic" collapse models has been
replaced by the conviction that galaxies are assembled from
smaller structures embedded within merging dark matter halos. If
gas cooling is inhibited by various feedback processes, then the
cosmic history of star formation will be spread over a wide range
in redshift (White \& Rees 1978, Baugh et al. 1998). This is
consistent with recent high redshift observations (Madau 1999,
Madau et al. 1996) although in quantitative detail the results
remains controversial because of the different techniques used and
the difficulties in quantifying the effect of dust obscuration.

The most commonly used diagnostic for determining the star
formation rate within high redshift objects relies on the
proportionality between the rest frame UV flux (1500 \AA\ --
2500 \AA) and the abundance of short-lived massive stars. Rates
derived from the population of Lyman break galaxies (Steidel et al. 
1999) can be compared with those from blue-selected intermediate
redshift field galaxies in the range $0.2<z<1$ (Lilly et al. 1996,
Cowie et al. 1999) as well as for observations of nearby galaxies
observed at 2000 \AA\ from FOCA, a balloon borne panoramic imaging
camera (Milliard et al. 1992, Treyer et al. 1998, Sullivan et al.
1999). The picture emerging from these surveys has the comoving
star formation rate rising by a large, but uncertain, factor over
0$<z<$1-2 and possibly declining more slowly at higher redshifts
(Steidel et al. 1999). Importantly, the only UV-based constraint on
the star formation rate in the region $1<z<2$, where the comoving
volume-averaged rate apparently peaks, comes from photometric
redshift data whose precision is unclear (Connolly et al. 1997).

An independent probe of star formation is based on nebular
emission lines measurements. Gallego et al. (1997) undertook an
objective prism survey of nearby galaxies and used the H$\alpha$
emission line luminosity density to infer a local star formation
rate assuming the line is produced by photoionisation of nebular
gas by the UV continuum of massive stars (Kennicutt 1998).
However, for $z>$ 0.4, H$\alpha$ is shifted beyond reach of most
optical spectrographs and searches must be undertaken in the near
infrared where detector technology is still rather primitive.
Glazebrook et al. (1999) compared estimates of the star formation
rate for a small sample of $z\simeq$1 galaxies using both CGS-4
spectroscopy of H$\alpha$ and the rest-frame UV continuum. They
found significant discrepancies which they attributed to both dust
extinction and possible timescale variations in the star formation
activity. However, the signal to noise of their infrared spectra
precluded a rigorous comparison.

[O {\sc ii}] is a somewhat less accurate tracer of star formation
although available in the optical regime to much higher redshifts
than H$\alpha$ (Cowie et al. 1995, Guzm\'an et al. 1997). However,
both emission line indicators are prone to uncertainties from the
unknown extinction which inhibits the escape of Lyman limit
photons (Flores et al. 1999). Recent observations at high $z$ in
the submillimeter band have emphasised these problems and revealed
a potentially separate population of dusty galaxies undergoing
intense star formation (Blain et al. 1999, Hughes et al. 1998,
Barger et al. 1998). Comoving star formation rates derived from
these sub-mm sources are also much higher than those found through
the UV continuum methods although uncertainties remain in
estimating the bolometric far-infrared flux solely from the sub-mm
detections and dust heating may, in many cases, arise from
non-thermal UV sources rather than massive stars. Indeed, recent
work suggests little disparity between the sub-mm and the Lyman
break populations when the various uncertainties are taken into
account (Steidel et al., in preparation).

In this paper we explore the possibility of providing new
constraints on the UV-based star formation rate by utilising deep
lensed images of background galaxies viewed through a
well-understood massive cluster. The bonus of using a massive
cluster is the significant magnification of those sources viewed
through it. The drawback, of course, is that to estimate a
comoving star formation rate requires a knowledge of both the
redshifts of individual star-forming galaxies (the faint arcs in
this case), as well as a good understanding of the survey
completeness and lensing mass model, for a detailed study of the
magnification variation within the field of view. 

Deep spectroscopic surveys of faint arcs have already demonstrated
that most lie between redshifts $z\simeq 0.6$ and 2 and several cases
exist of more distant, highly-magnified sources (Mellier et al. 1991,
Pell\'o et al. 1999, and other references in B\'ezecourt et al. 1999).
Independent redshift estimates can also be made utilising the fact
that the image distortion becomes more significant for a source at
higher redshift. By measuring the shape of an arclet it is possible,
in the context of a well-constrained mass model for the lensing
cluster, to infer a probability distribution for the source redshift by
searching for the most circular source corresponding to the observed
image (Kneib et al. 1996, Ebbels et al. 1998).  Finally multicolour
imaging offers a third way of estimating redshift. This has been most
successful through the Lyman break surveys ($z>2.3$, Steidel et
al. 1999) and via photometric redshift surveys up to $z\simeq 1$ (Hogg
et al. 1998). However, useful constraints in the interval $1<z<2$
appears to require optical and near infrared data and the predictions,
although promising, need to be verified spectroscopically (Pell\'o et
al. 1998).

In this paper, we analyse a deep Hubble Space Telescope (HST) Wide
Field Planetary Camera 2 (WFPC2) ultraviolet image (F336W) of the
rich cluster Abell 370 ($z=0.37$) in order to infer the abundance,
redshift distribution and intrinsic properties of lensed star
forming objects with $z<3$ $\footnote{The redshift limit arises
from considerations of the location of the Lyman limit}$. UV
imaging is a very efficient way to select background star forming
objects as cluster ellipticals disappear in this waveband.
Moreover, by comparing the detection of these UV arclets with a
deeper sample located in a R (F675W) HST image, it may be possible
to infer redshift constraints from both the presence or otherwise
of a Lyman break inbetween the two bands.

Section 2 presents the new HST data and discusses the reduction
procedures used to construct a photometric catalogue. In Section
3, we discuss the expected number counts and redshift distribution
for UV-selected arclets in the specific case of Abell 370 using
the evolutionary code and lensing mass model discussed in
B\'ezecourt et al. (1999, hereafter Paper I). The predictions are
compared with the observational data. In Section 4, further
constraints on the redshifts of individual arclets are derived
using a variant of the lensing-inversion method. The results
enable us to derive the redshift-dependent comoving star formation
rate in Section 5 as well as to compare our estimates with other
UV-based values. Our conclusions are summarised in Section 6.
Throughout the paper, we adopt a Hubble constant of H$_0 = 50 \,
{\rm km \,s}^{-1} {\rm Mpc}^{-1}$, $\Omega_0= 1.0$ and
$\Omega_{\Lambda} = 0$.

\section{Ultraviolet Photometry}
\begin{figure*}
\centerline{\psfig{figure=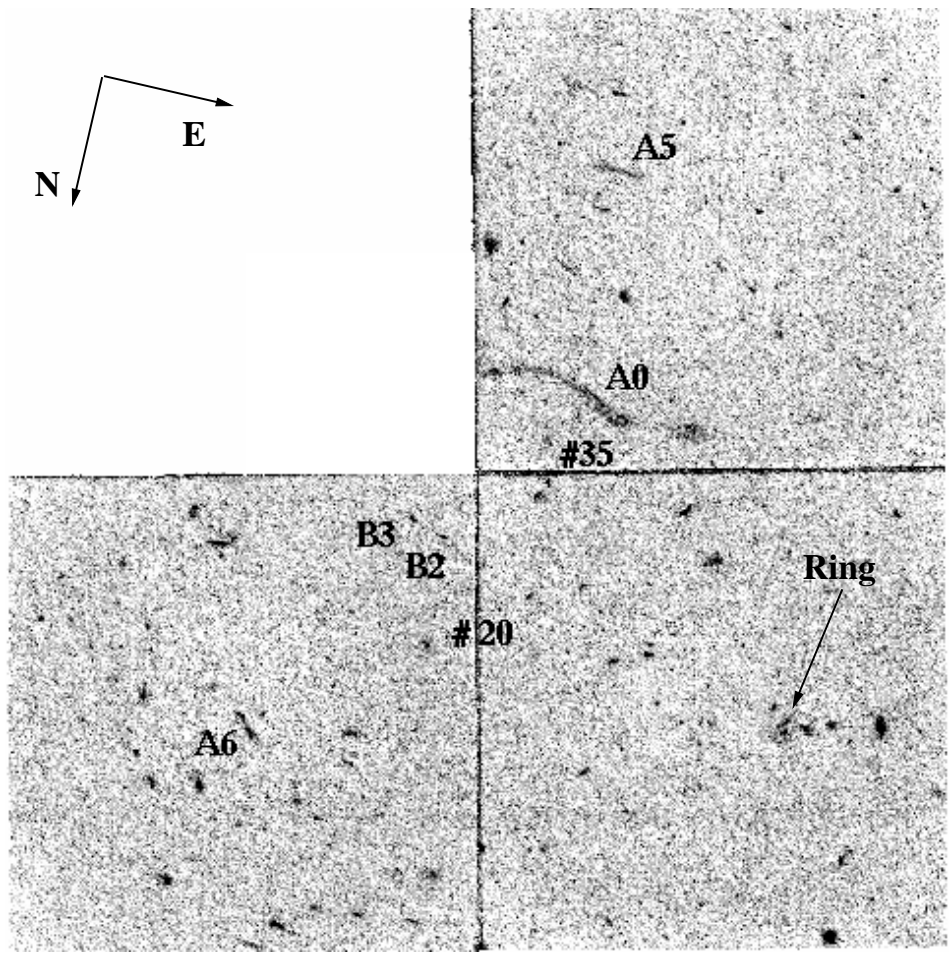,width=0.7\textwidth,angle=-90}}
\caption{Hubble Space Telescope Wide Field Planetary Camera 2
F336W image of Abell 370 ($z=0.37$). Most of the arclets
identified in the ground-based study of Fort et al. (1988) are
revealed as well as the spectacular ring galaxy identified by Soucail
et al. (1999). By contrast most of the cluster members are effectively
transparent at this wavelength. The shear field can be readily
seen and is perpendicular to the axis defined by most luminous
members \#20 and \#35.} \label{ima_F336W}
\end{figure*}

The deep ultraviolet image of Abell 370 which forms the basis of
this study was obtained with the Hubble Space Telescope (HST) Wide
Field Planetary Camera 2 (WFPC2) [ID: 5709, P.I.: J.M. Deharveng]. It 
was originally taken to study the ultraviolet spectral properties
of member elliptical galaxies. For this purpose, exposures were
taken with the F336W filter ($T_{exp}$ = 16.2 ksec) and the F225W
filter ($T_{exp}$ = 11.6 ksec). No objects were detected in the
F225W image and thus only the F336W image will be discussed in
this paper.

The F336W data was reduced using standard procedures within the
IRAF/STSDAS packages. One difficulty that arose was the fact that
the individual WFPC2 frames had been binned directly on the
telescope. Accordingly, the cosmic ray rejection was slightly
sub-optimal. The final reduced F336W image is displayed in Figure
\ref{ima_F336W}. Indeed, most of the cluster members are invisible
at 336 nm, contrary perhaps to expectations, and the only visible
sources appear to be background, lensed star-forming galaxies.
Most of the candidate arclets identified in the ground-based study
of Fort et al. (1988) are visible in the F336W image. These UV images
appear to reveal a significant shear field perpendicular to the
axis defined by the two dominant galaxies.

Of these 37 candidates arcs 23 lie inside the F675W field and 
22 are detected (not all with $a/b>2$). Although the reverse comparison
could potentially be useful in locating candidate F336W `dropouts',
unfortunately the UV image is not deep enough to realise this aim.
Of 81 candidate arclets identified by B\'ezecourt et al (1999) in the F675W,
55 are in the F336W field and thus half remain undetected.

Photometry of the detected images was performed using the
Sextractor package (\cite{bertin96}) calibrated using zero-points
given by Holtzmann et al (1995). Image detection was performed
assuming a minimum of 12 contiguous WFPC2 pixels (48 unbinned
pixels) above 2$\sigma$ of the local sky level in order to limit
the number of spurious objects appearing with the high level of
background noise. Although this area limit is somewhat larger than
that adopted for the F675W image discussed in Paper I, it is
adequate for detecting distorted arclets whose typical areas lie
well above this threshold. The adopted 1$\sigma$ surface
brightness detection limit is $U_{336W}=22.1$ mag arcsec$^{-2}$
and, from the magnitude histogram obtained from the complete
catalogue, we estimate a completeness limit of $U_{336W}=23$. As a
working hypothesis we will define an arclet as an image whose axis
ratio ($a/b$) is larger than 2. The number of arclets with
$U_{336W}<23.5$ which constitutes our working sample is 37.

\section{Model Predictions}
\subsection{Number-magnitude and number-redshift distributions}

B\'ezecourt et al. (1998) developed techniques capable of
predicting both the number-magnitude $N(m)$ and the number-redshift
$N(z)$ distributions of lensed arclets by coupling standard models of
galaxy evolution valid in unlensed fields (Bruzual \& Charlot 1993 and
\cite{pozzetti96}) with the lensing magnifications expected for
clusters with detailed mass models.  The galaxy evolution models are
tested by reproducing, so far as is known, $N(m,z)$ for field galaxies
observed in various wavebands. Briefly, the models assume 4 population
types, a local luminosity function and two generic models for the
evolution depending on the cosmological parameters. The $\Omega$=0
model assumes number-conserving pure luminosity evolution of the form
advocated by Pozetti et al. (1996), whereas for $\Omega$=1, number
evolution is included. Together with a knowledge of the cluster mass
distribution, derived independently from lensing constraints based on
series of multiple images with either direct
spectroscopy or photometric estimates (c.f. Kneib et al. 1996), the
unlensed source counts and redshift distribution can be modified
accordingly for comparison with the observations.

This method was applied to the cluster Abell 370 in Paper I
(B\'ezecourt et al. 1999) using an improved lens model constrained
by lensed images visible in a HST F675W image (ID: 6003, P.I. :
Saglia) and new ground-based spectroscopy. The cluster mass
distribution was modelled using a large scale component augmented
with smaller scale mass contributions representing the brightest
cluster members scaled according to their luminosity
(\cite{kneib96}). The intrinsic ellipticity distribution of the
faint sources was included in the model predictions. The predicted
counts of arclets with $a/b>$2 agree well with those derived
directly from the F675W image. Moreover, the models imply a
significant fraction of arclets with $R<23.5$ have redshifts
$z\simeq 1-5$ in contrast to that expected for an unlensed field
survey.

\begin{figure}
\psfig{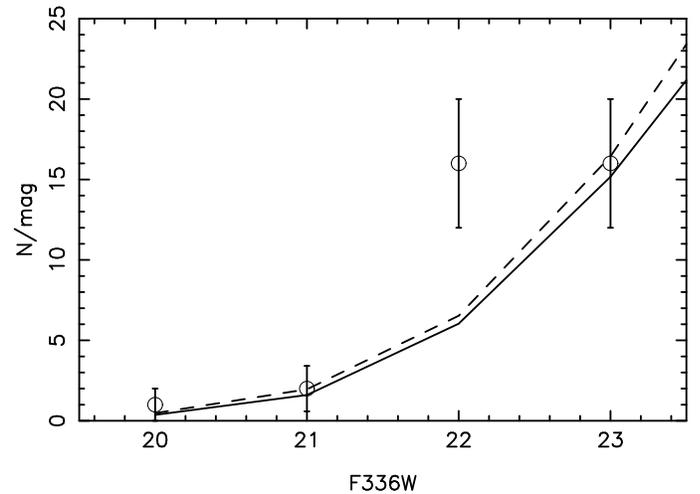}
\caption{The observed number magnitude counts ($\circ$) of arclets
observed with $U_{336W}<23.5$ and an axis ratio $a/b>2$ in Abell
370 in the F336W filter compared to models incorporating standard
evolutionary changes and gravitational magnifications determined
statistically through the adopted cluster mass model (see text and
Paper I for technical details). The solid line refers to the model
based on $\Omega_0=1$ and the dashed line to that for which
$\Omega_0=0$.} \label{histo_F336W}
\end{figure}

We now extend the Abell 370 model developed in Paper I in order to
predict the $N(m,z)$ distributions expected for arclets selected at
ultraviolet wavelengths. The number magnitude distribution for F336W
arclets observed with $a/b>$2 is compared in Figure
\ref{histo_F336W} to model predictions according to the precepts
discussed in Paper I for $\Omega_0=0$ and 1 as introduced above.  It is
important to realise that this comparison, whilst useful, is approximate
in a number of respects. Firstly, the model predictions do not account
for any surface brightness threshold inherent in the actual imaging data
and are, instead, based on integrated magnitudes only. Secondly, the
lensing magnification is determined statistically assuming a uniform
distribution of background sources; for a small sample it is likely
there will be significant deviations from the prediction depending on
the interplay of source clustering and the large magnification
variations across the lens (we will return to this point in
\S3.2). Finally, the absolute normalisation of the local UV galaxy
luminosity function, essential for an accurate comparison, is poorly
known. Armand \& Milliard (1994) and Fioc \& Rocca--Volmerange (1999)
have proposed an additional population of star-forming sources in order
to account for the FOCA balloon-borne counts of galaxies selected at 2000
\AA . The excess population inferred from the FOCA source counts over and
above optically-derived estimates amounts to a factor of $\simeq 2$ (Armand
\& Milliard 1994, Treyer et al. 1998). Such an excess population would
significantly improve the comparison in Figure \ref{histo_F336W}
making it reasonable considering the uncertainties.

\begin{figure}
\psfig{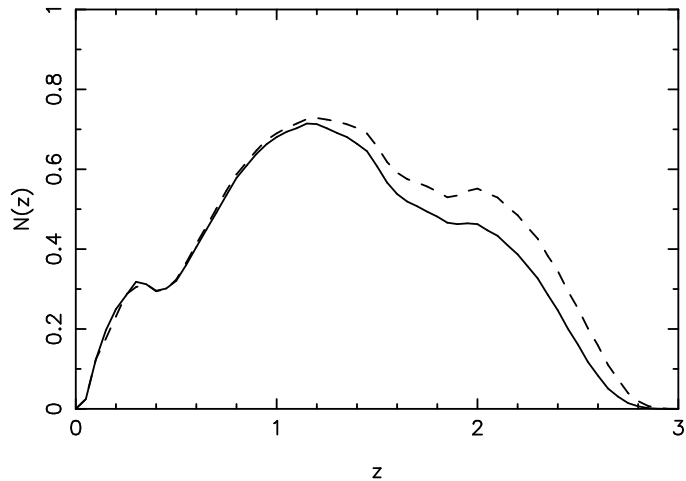}
\caption{The expected redshift distribution for UV-selected
arclets viewed through Abell 370 according to the selection
criteria $U_{336W}<23.5$ and axis ratio $a/b>2$. The solid line
corresponds to $\Omega_0=1$ and the dashed line to $\Omega_0=0$.}
\label{redshift_336}
\end{figure}

The predicted arclet redshift distribution based on the
statistical lensing expected through the Abell 370 mass model is
shown in Figure \ref{redshift_336} for a magnitude limit of
$U_{336W}=23.5$ and an axis ratio limit $a/b>2$.
The effect of gravitational magnification is clearly visible;
below $z\simeq 0.4$ the population is entirely due to unlensed
field galaxies. Our selection criteria appear to draw sources
primarily with $z\simeq 0.6-2.2$. The decline in numbers at higher
redshift arises from the entry of the discontinuity around
Ly$\alpha$ and later the Lyman break through the F336W filter. A
significant advantage of the UV HST image is therefore its ability
to select sources in a redshift range which is largely
inaccessible via conventional optical spectroscopy. Contrary to
the redshift distribution in the R-band (Paper I), there is no
strong difference between the two cosmological cases because the 
redshift range stops at $\simeq$2.2.

It is important to realise that no internal extinction has been
assumed for any of the faint UV sources. Dust extinction is
critically important since rest-frame wavelengths of less than
$\simeq 1500$ \AA\ are implied for $z>1$. Extinction would reduce
the number of arclets observed and, if the effect was more
pronounced for high redshift sources, the observed peak of the
redshift distribution would shift to lower redshifts. In this
respect, modulo the uncertainties in the modelling predictions,
our data may provide a sensitive probe of the dust
distribution at high redshift.

\subsection{The spatial distribution of arclets}

\begin{figure}
\psfig{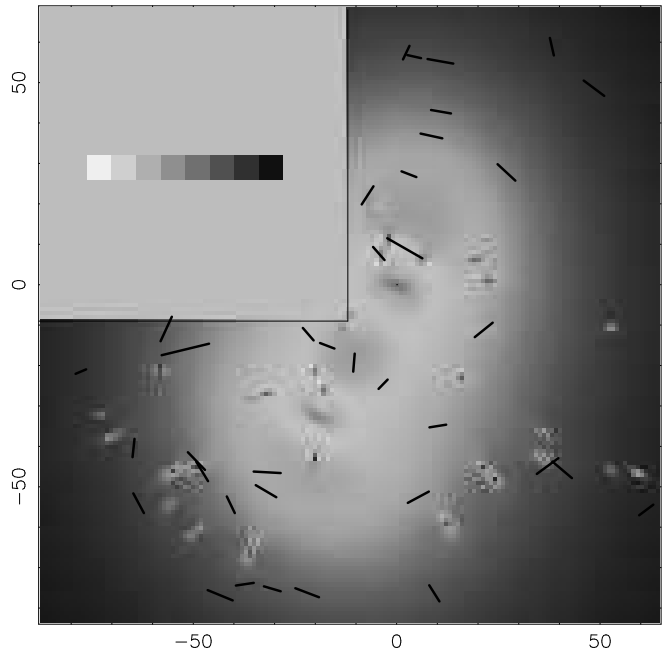}
\caption{The surface density of arclets with $U_{336W}<$23.5 and
$a/b>2$ in Abell 370 assuming $\Omega_0=0$. The grey scale
indicates the ratio $N$(lensed)/$N$(field) and runs from 0 (white)
to 1 (black). Arclets detected on the HST F336W image are
indicated by ticks whose lengths and orientations are consistent
with their observed deformations.}
\label{fig-cartes_336}
\end{figure}

We now return to the point raised in $\S$3.1 and consider whether
the spatial distribution of arclets within the HST image is
consistent with that expected for the adopted mass model given the
model parameters and detection limits. Similar comparisons were
made for Abell 2218 in B\'ezecourt et al. (1998) where the
methodology is discussed in detail.

We now apply this procedure to the UV HST image of Abell 370 in
order to see where the arclets are expected to lie. The resulting
density map is compared to the distribution of the detected
arclets in Figure \ref{fig-cartes_336}. The agreement is fairly
reasonable. The giant arc A0 is located close to a fairly high
density region and few arclets are found in the low probability
areas. The lowest density occurs in the area enclosed by critical
lines between $z\simeq 0.5$ and $z\simeq 3$ where there is a
characteristic depletion in the number counts arising from the
so-called magnification bias (Broadhurst 1995, Fort et al. 1996).

Figure \ref{fig-cartes_336} can be used to improve the likelihood
that an arclet candidate is a true lensed source as well as
aiding spectroscopic follow-up surveys. Elongated objects that
occur in low probability areas can be rejected in a first step and
more attention can be devoted to regions where the probability
density that an elongated object is lensed is higher, although the
contamination by elongated cluster members will always remain
non-zero. 
Such images will be useful tools for optimal selection of arclets in 
spectroscopic surveys.

\section{Estimating the arclet redshifts}

As the strength of gravitational distortion for a particular
arclet depends not only on the lensing potential and geometrical
alignment but also on the redshifts of the source and lens, it is
possible from the deformation properties of a given arclet to
derive a probabilistic estimate of the source redshift assuming
some distribution for its intrinsic size and shape. The latter
can be provided by considering the properties of faint field
galaxies observed by HST in unlensed areas. This method was
developed by Kneib et al. (1996) and applied in the context of HST
data for the cluster Abell 2218. Arclet redshifts, predicted on
the basis of a mass model, were later verified successfully
through spectroscopic follow-up (\cite{ebbels98}).

In the previous section, we predicted the angular distribution of
lensed images from F336W-selected source counts modified by the
lensing mass model for Abell 370 (Figure \ref{fig-cartes_336}).
Implicit in this calculation was the redshift distribution
expected, statistically, for the source population.

Similarly, it is also possible to compute {\it for each arclet}
the redshift probability distribution consistent with a lensed
image whose orientation and axis ratio matches the observed value.
The weighted average of this distribution is then assigned to the
arclet as its {\em most probable redshift}. The only constraints
involved are the deformation vector $\tau$ (c.f. Kneib et al 1996)
of the arclet and the limiting magnitude of the sample. This
redshift estimate corresponds to an improvement of the method
adopted by Kneib et al., as the redshift probability distribution
derived from the lensing distorsions is now weighted by that
expected for the selected filter and sample magnitude limit. 
The fact that the arclets are brighter than the limiting magnitude 
in filter F336W give constraints on their redshift which are 
then combined with the purely lensing constraints to give a better
redshift estimate. 

\begin{table*}
\caption{Estimated arclet redshifts for the $a/b>2$ F336W sample
in Abell 370$^{\ast}$}
\label{table_z}
\begin{flushleft}
\begin{tabular}{rrrccccccc}
\hline\noalign{\smallskip}
object&X&Y&a/b&U$_{336W}$&$z_{est}$&$\sigma_z$&SFR(1500 \AA )&SFR(2000 
\AA )&other\\
 &($''$)&($''$)& & & & &(M$_\odot$/yr)&(M$_\odot$/yr)&identification\\
\noalign{\smallskip}
\hline\noalign{\smallskip}
 1& --43.4& --76.8& 2.84& 21.16& 1.80& 0.39&31.0&19.0\\
 2& --22.0& --76.2& 4.46& 22.22& 1.71& 0.31&6.0&3.7\\
 3&    9.3& --76.4& 2.44& 22.40& 0.34& 0.21&1.1&0.69\\
 4& --30.6& --75.2& 2.67& 22.56& 1.36& 0.34&3.2&2.0\\
 5& --37.3& --74.1& 2.30& 22.17& 0.53& 0.21&2.3&1.4\\
 6&   61.3& --55.7& 2.66& 23.14& 1.65& 0.44&4.6&2.8\\
 7& --40.8& --54.4& 3.16& 23.25& 0.72& 0.13&0.82&0.50\\
 8& --63.4& --54.1& 2.06& 21.89& 1.17& 0.37&6.4&3.9\\
 9&    5.3& --52.6& 3.31& 22.37& 1.03& 0.15&1.3&0.81\\
10& --32.1& --51.1& 4.18& 23.34& \underline{\it1.4} & 0.07&0.88&0.54& A1\\
11& --31.8& --46.4& 2.57& 21.77& 0.41& 0.10&2.1&1.3\\
12& --47.8& --46.2& 2.38& 21.99& 1.23& 0.32&3.0&1.8\\
13&   40.7& --45.8& 2.25& 21.76& 0.37& 0.25&2.2&1.4\\
14&   37.1& --44.9& 3.70& 22.51& 1.34& 0.30&3.0&1.8\\
15& --49.2& --43.6& 2.73& 22.29& 0.69& 0.09&1.6&1.0\\
16& --64.7& --40.4& 2.24& 21.69& 0.92& 0.34&5.7&3.5\\
17&   10.1& --35.0& 2.06& 22.44& 0.61& 0.18&1.2&0.70\\
18&  --3.3& --24.6& 2.10& 22.91& 0.28& 0.11&0.49&0.30\\
19& --77.6& --21.5& 2.23& 23.11& 0.43& 0.36&0.85&0.52\\
20& --10.5& --19.3& 4.37& 23.32& \underline{\it1.3} & 0.10&3.7&2.3& E2\\
21& --17.1& --15.1& 3.14& 22.89& \underline{\it0.806}& 0.002&0.40&0.25& B2\\
22& --51.9& --16.0& 3.65& 20.99& 0.31& 0.16&3.5&2.2\\
23& --21.7& --12.2& 2.39& 22.46& \underline{\it0.806}& 0.002&0.76&0.46& B3\\
24& --56.6& --11.0& 2.27& 21.63& 1.06& 0.33&5.7&3.5\\
25&   21.4& --11.2& 2.87& 21.38& 0.72& 0.11&2.6&1.6\\
26&  --4.4&    7.7& 4.43& 23.53& 0.25& 0.09&0.22&0.13\\
27&    2.0&    9.0& 10.0& 20.07& \underline{\it0.725}& 0.001&2.3&1.4& A0\\
28&  --7.2&   22.1& 3.40& 22.60& 0.26& 0.10&0.55&0.33\\
29&   10.8&   55.3& 4.10& 22.38& 1.91& 0.48&6.5&4.0\\
30&    4.3&   56.4& 2.29& 23.48& 1.17& 0.35&1.3&0.78\\
31&    2.4&   57.4& 2.05& 23.20& 0.35& 0.23&0.57&0.35\\
32&   38.1&   58.9& 3.90& 23.46& 0.35& 0.16&0.45&0.28\\
33&    3.0&   27.3& 2.48& 23.35& 0.60& 0.09&0.41&0.25\\
34&   27.0&   27.8& 2.60& 22.86& 0.97& 0.23&1.3&0.79\\
35&    8.5&   36.8& 2.82& 22.64& 0.84& 0.14&1.3&0.78\\
36&   10.9&   42.8&  5.0& 21.58& \underline{\it1.3} & 0.26&4.2&2.6& A5\\
37&   48.5&   48.6& 2.35& 22.23& 1.81& 0.53&14.3&8.8\\
\noalign{\smallskip}
\hline
\end{tabular}
\end{flushleft}

$^{\ast}$X,Y positions are defined with respect to galaxy \# 20.

The redshifts for A0, A1,A5,B2, B3 and E2 are taken from Soucail
et al. (1988), B\'ezecourt et al. (1999) and Mellier et al.
(1991).

Star formation rates were computed from the luminosity at 2000
\AA\ according to the Donas et al. (1990) calibration and at 1500 \AA\ 
according to the Madau et al. (1998) calibration

\end{table*}

\begin{figure}
\psfig{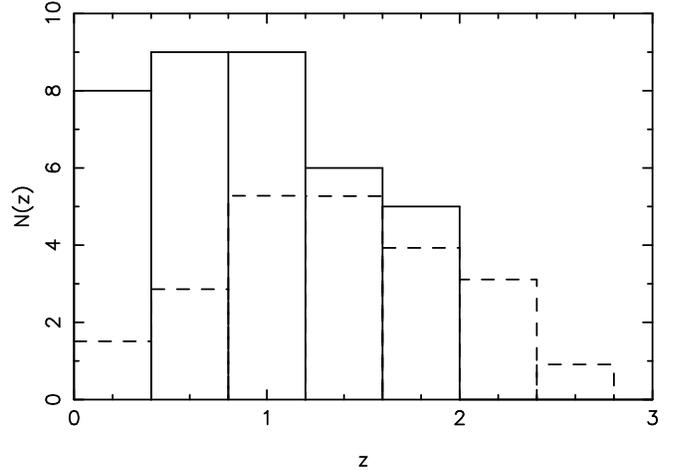}
\caption{A comparison of the redshift distribution of UV arclets
in Abell 370 inferred from lensing inversions (solid line) and the
the model in the HST field presented as a solid line in figure
\ref{redshift_336} (dashed line). The excess at low redshift may
represent an inadequacy in the absolute normalisation adopted for
the local UV luminosity functions (Treyer et al. 1998).}
\label{z_uv_model}
\end{figure}

The lens inversion redshifts range from 0.3 to 1.9 for the objects
in the sample (Table \ref{table_z}). We compare these estimates
with the redshift distribution derived via the models discussed in
Section 3 in Figure \ref{z_uv_model}. As found by Ebbels et al.
(1998), redshift estimates are less secure where the lensing power
is small which may explain some of the differences seen. Even
allowing for the difference in absolute numbers discussed earlier
(c.f. Figure 2), at low redshift, the lensing method predicts a
significant excess of galaxies over the model predictions.  This
is, in part, due to the excess UV counts discussed earlier. It
would seem most of this excess lies at low redshift as discussed
by Treyer et al. (1998). There is also a discrepancy beyond
$z\simeq 2$ where the model predicts a tail not seen in the
lensing-inferred distribution. Again, as discussed earlier, there
are many reasons why the modelled $N(z)$ overestimates the mean
redshift of the F336W population. Considering the uncertainties,
the distribution of arclet redshifts in the important region
$0.5<z<2$ is reasonably well matched.

\section{Cosmic star formation rates}

The redshift distribution of UV-selected arclets discussed in
Section 4 enables us to estimate the volume-averaged star
formation rate as a function of redshift in an important interval
$0.5<z<2$ independently of methods used by other workers. By
selecting arclets at F336W, our observations span a much shorter
rest-frame wavelength range ($\lambda\lambda$ 1200--2200 \AA\
typically) than in other studies of high redshift galaxies and
consequently are more sensitive to both star formation activity
and dust extinction.

To estimate the star formation rate, we adopt a proportionality
between the UV ionizing flux emitted by massive and short-lived
stars which has been calibrated by several authors. Donas et al.
(1990) used direct UV observations of local galaxies at 2000 \AA\
and give the relation
\begin{equation}
\label{eq-donas}
SFR \; (M_{\odot}/yr)=L_{2000} \ \left(h_{50}^{-2} \ {\rm erg\,
s}^{-1} {\rm \AA}^{-1}
\right) / 7.58\times 10^{39}
\end{equation}
whereas Madau et al. (1998) use the scaling relation at 1500 \AA\
\begin{equation}
SFR \; (M_{\odot}/yr)=L_{1500} \ \left(h_{50}^{-2} \ {\rm erg\,
s}^{-1} {\rm Hz}^{-1} \right) / C
\end{equation}
where $C$ is a conversion factor which depends on the choice of
the initial mass function (IMF): $C = 8 \times 10^{27}$ for a
Salpeter IMF and $C = 3.5 \times 10^{27}$ for a Scalo IMF. Note
that there is a factor 2 of difference in this conversion
depending on the choice of the IMF. To be consistent with other
authors, we will consider a Salpeter IMF in the following and
adopt $\Omega_0=1$.

\subsection{Methodology}

Using the redshift assigned to each arclet, the rest-frame
luminosity $L_{\lambda}$ (in erg s$^{-1}$ \AA$^{-1}$) at a given
wavelength (1500 or 2000 \AA\ for our purpose) becomes:
\begin{equation}
L_{\lambda}=  k_\lambda \, (1+z) \times {1 \over A(z)} \,
10^{-0.4(m_{F336W}+21.1)} \times 4 \pi \, d^2_L (z)
\end{equation}
where $k_\lambda$ is the differential k-correction necessary to
convert the rest-frame flux at ${3360 / (1+z)}$ \AA\ to
$\lambda=1500$ \AA\ or 2000 \AA :
\[
k_\lambda = {f_{\lambda} \over f_{3360 / (1+z)}}
\]

The differential k-correction is small, and consequently not too
model dependent because, for $z\simeq 0.5-2$, the rest-frame
wavelength of the F336W filter always lies close to the selected
UV wavelength. The correction was determined using the model
spectral energy distribution of a typical star-forming Sd galaxy.

$d_L (z)$ is the luminosity distance at redshift $z$ and $A(z)$ is
the magnification factor of the arclet derived from the mass model
for Abell 370.

If we remove from the sample given in Table 1 all arclets whose
lensing-inferred redshift indicates $z<0.4$ (i.e. the inversion
method suggests they are either cluster members or foreground
galaxies) as well as one arc which is a double image (B2/B3), we
have 28 arclets. This residual sample can be divided into two bins
with similar numbers:  13 arclets in the redshift interval
$0.4<z<1$ (Bin 1) and 15 arclets in the interval $1<z<2$ (Bin 2).

We recall that our sample is limited to arclets with an axis ratio
larger than 2 in order to increase the probability that they are
background sources. This means that the galaxies we study are
drawn from a volume which is not necessarily that defined simply
according to a fixed F336W magnitude limit. In effect the volume
probed corresponds to the integral with redshift of the area in
the source plane (at a given redshift) which magnifies arclets to
an axis ratio $a/b>2$, $S(z | b/a > 2)$: 
\[ {\cal V} [z_1, z_2] = \frac{c}{H_0} \ 
\int_{z_1}^{z_2} \frac{dS(z | b/a > 2)}{4 \pi} 
\ \frac{d_L^2 (z)}{(1 + z)^3} \ \sqrt{2 q_0 z + 1} \ dz
\]
The corresponding volumes are respectively $\cal V$[0.4,1] = 1130
$h_{50}^{-3}$ Mpc$^3$ and $\cal V$[1,2] = 2910 $h_{50}^{-3}$ Mpc$^3$.

\subsection{Star formation rates}

\begin{figure}
\centerline{ \psfig{figure=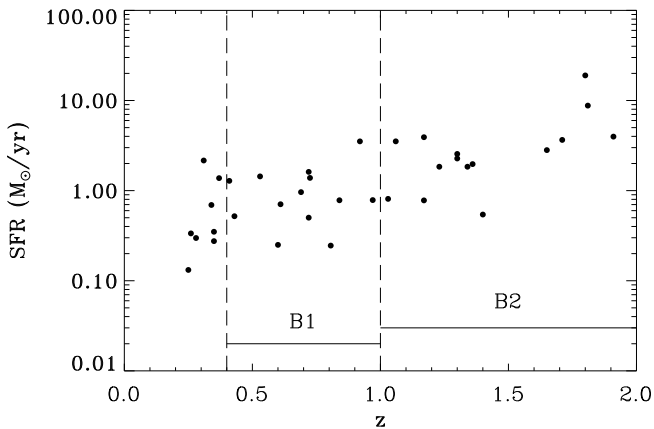,width=0.5\textwidth} }
\caption{Star formation rate (in units of $h_{50}^{-2}$) versus
redshift inferred from F336W fluxes for arclets detected with an
axis ratio greater than 2 in the HST F336W image.}
\label{sfr_objets}
\end{figure}

Given the above procedures we can infer the mean star formation
rates for individual arclets and compare these with those derived
for other high redshift sources. Figure \ref{sfr_objets} shows the
trend with redshift using the Donas et al. (1990) relation to
convert the 2000 \AA\ luminosity. The distribution shows a modest
increase of the SFR across the two broad redshift bins whose
significance is discussed below.

The mean star formation rate per object for our arclet sample is
$\simeq$ 2.5 $h_{50}^{-2}$ M$_{\odot}$/yr. Using the Madau et al.
relation would increase this to $\simeq$ 4 $h_{50}^{-2}$
M$_{\odot}$/yr (Salpeter IMF) or 9.3 $h_{50}^{-2}$ M$_{\odot}$/yr
(Scalo IMF). Such star formation rates are consistent with those
inferred for arcs seen in other clusters (e.g. in Abell 2390
\cite{bezecourt97}) as well as for other UV-selected sources at
higher redshift (Steidel et al. 1999).

In order to address the possible increase in mean star formation
rate with redshift apparent in Figure 6 as well as to compute the
redshift-dependent comoving star formation rate, we need to
estimate the incompleteness correction in both bins. The most
direct way, following Lilly et al (1996) and Connolly et al
(1997), is to compare our equivalent luminosity function at
2000\AA\ with the one spectroscopically determined at low redshift
($z<0.3$) by Treyer et al. (1998) and correct for that fraction of
sources fainter than the limit appropriate at a given redshift.
Adopting the Treyer et al. 2000\AA\ Schechter luminosity function
whose faint end slope is $\alpha=-1.6$ we derived correction
factors of 1.6 for Bin 1 and 2.1 for Bin 2. The uncertainty
depends primarily on the slope of the faint end of the luminosity
function which is, of course, not yet constrained at the redshifts
in question. Had we adopted Lilly et al.'s slope of $\alpha = -1.3$
the correction factors would have been reduced to 1.2 for Bin 1
and 1.4 for Bin 2. This gives some measure of the uncertainties
involved.

\begin{figure}
\centerline{
\psfig{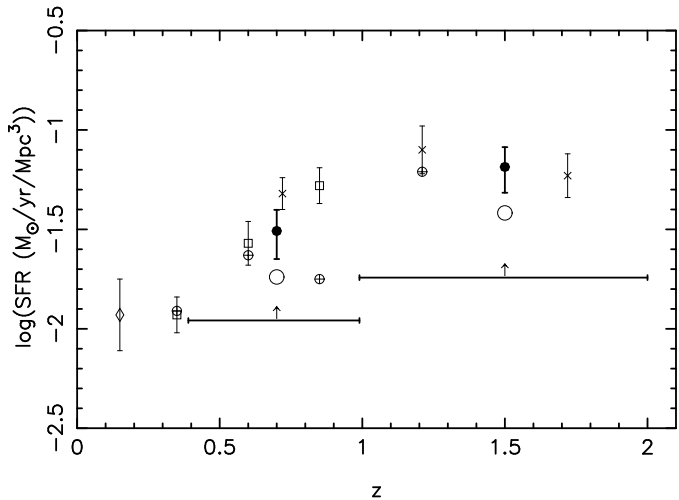} }

\caption{Comoving star formation densities for the Abell 370
arclet sample derived according to the UV conversions relations of
Donas et al. ($\circ$) and Madau et al. ($\bullet$) using a
Salpeter IMF. For clarity, the error bar based on statisical
uncertainties is only shown on the point based on the Madau et
al. (1998) 
conversion. The width of the redshift bins is indicated for the
lower limit on the SFR derived with the scaling of Donas et al.
(1990) without correction. Other UV-based determinations include:
$\diamondsuit$: Sullivan et al. (1999), 
$\times$: Connolly et al. (1997),
$\Box$: Lilly et al. (1996), $\oplus$: Cowie et al. (1999).}
\label{sfr_volume}
\end{figure}

Applying this correction, we show the comoving star formation
rate, uncorrected for possible extinction effects, in Figure
\ref{sfr_volume} for our two redshift intervals. The inevitably
larger incompleteness correction for Bin 2 strengthens the
increase of SFR with redshift seen in Figure 6 although the
statistical uncertainties clearly are large. For convenience we
show the effect of using both the Donas et al. (1990) and Madau et
al. (1998) conversions (which differ by a factor of 1.6). Other
UV-based measurements are shown in order to allow a consistent
comparison between different samples with roughly similar
extinction bias.

The absolute value of the star formation density remains unclear
in all the surveys compared because of the high sensitivity of the
UV to SFR conversion to the initial mass function and because of
the damaging effect of even modest amounts of dust extinction. For
an extinction of 1.2 magnitudes at 2000 \AA\ (as argued by Buat
\& Burgarella (1998) on the basis of a sample of starburst
galaxies) the corrected rates would be twice as high. Evidence for
significant extinction has also been given by Fanelli et al. (1996),
who discuss a discrepancy of a factor $\simeq $ 2.5--9 between rates
derived in the UV (1500 \AA) and the far-infrared. Yan et al. (1999)
also find a discrepancy factor of 3 between H$\alpha$-based rates
and those from the continuum at 2800\AA\ in the redshift range
$0.7<z<1.9$. 

Glazebrook et al. (1999) argue that part of the discrepancy between
$H\alpha$-based star formation rates and those utilising the UV
continuum may arise, in small samples, from the different
timescales of main sequence evolution involved. Sullivan et al
(1999) undertook a detailed comparison for a large sample of
UV-emitting galaxies with $z<0.3$ for which $H\alpha$ measures
were simultaneously available. A significant fraction of the
dispersion in the UV-H$\alpha$ luminosity correlation arises from
such effects. In principle, dust extinction could be calibrated
for many of the arclets observed here through multi-object
near-infrared spectroscopy of the Balmer lines which may become
practical shortly on large telescopes. Gravitationally lensed arcs
represent a highly practical alternative to normal field galaxies
for such probes of the intrinsic star formation rate because of
their high surface density and the fact that lensing constrains
their redshift in addition to photometric techniques useful for
field galaxies.

\section{Conclusions}

Using the combination of models for galaxy evolution introduced in
B\'ezecourt et al (1999, PaperI) and a mass distribution for
cluster Abell 370 constrained by recent lensing studies, we
discuss the statistical properties expected for a lensed sample of
$\simeq 30$ faint arclets in a deep Hubble Space Telescope image
taken using the F336W filter with the Wide Field Planetary Camera
2. By comparing predicted source counts and their redshift
distribution with the spatial distribution and numbers of observed
arclets in the HST image, we conclude that the bulk of the UV
sources seen are gravitationally lensed and lie in the redshift
range 0.5--2. As such they offer a novel probe of the star
formation history in this relatively unexplored redshift range.

Using the observed shape of each arclets in the context of the
adopted mass model, we derive its most likely redshift using a
modified version of the lensing inversion technique developed by
Kneib et al. (1996). We compare the inversion-based redshift
distribution with that predicted on the basis of the evolutionary
models and find general agreement at high redshift although it
seems a significant fraction of UV sources represent an excess of
low redshift star forming sources consistent with the survey of
Treyer et al. (1998).

Assuming an incompleteness correction based on the steep UV
luminosity function observed for $z<0.3$, we derive star formation
densities consistent with other UV-based estimates. The typical
arcs seen in our sample imply star formation rates, uncorrected
for internal extinction, of a few $M_{\odot}$ yr$^{-1}$.

The spectroscopic follow up of deeper UV-based arclet samples
would permit tighter constraints through comparisons of
H$\alpha$-based star formation rates and corrections for
extinction based on Balmer line ratios. The high surface density
of magnified sources seen through rich clusters with
well-constrained mass distributions offers significant advantages
for probing the star formation characteristics in the $1<z<2$
regime over traditional field surveys.

\acknowledgements We thank R. Pell\'o for fruitful discussions and
encouragement, in particular about photometric and lensing-based
redshift techniques. This research has been conducted under the
auspices of a European TMR network programme (Contract No.
ERBFMRX-CT97-0172) made possible via generous financial support
from the European Commission ({\tt
http://www.ast.cam.ac.uk/IoA/lensnet/}) and through support of the
Programme National de Cosmologie and CNRS.

\end{document}